\documentclass[twocolumn,showpacs,showkey,preprintnumbers,amsmath,amssymb,floatfix]{revtex4}

\usepackage{graphicx,epsfig}
\usepackage{dcolumn}
\usepackage{bm}

\begin{document}

\preprint{}

\title{Phonon dispersion and low energy anomaly in CaC$_6$.}

\author{Matteo d'Astuto}%
\email{matteo.dastuto@impmc.upmc.fr}
\author{Matteo Calandra}%
\author{Nedjma Bendiab}%
\altaffiliation{present address: Laboratoire de Spectrometrie
  Physique, Universit\'e Joseph Fourier -
140 Av. de la physique,
BP 87 - 38402 Saint Martin d'H\`eres}
\author{Genevi\`eve Loupias}%
\author{Francesco Mauri}%
\affiliation{Institut de Min\'eralogie et de Physique des
Milieux Condens\'es (IMPMC), Universit\'e Pierre et Marie Curie -
Paris 6, case 115, 4, place Jussieu, 75252 Paris cedex 05, France}
\altaffiliation{Institut de Min\'eralogie et de Physique des Milieux
Condens\'es (IMPMC), CNRS UMR 7590, Campus Boucicaut, 140 rue de
Lourmel, 75015 Paris, France}

\author{Shuyun Zhou} %
\author{Jeff Graf} %
\author{Alessandra Lanzara}%
\affiliation{Materials Sciences Division, Lawrence Berkeley National
  Laboratory, Berkeley, CA 94720, USA} \affiliation{Department of
  Physics, University of California Berkeley, CA 94720, USA}

\author{Nicolas Emery}%
\altaffiliation{present address: University of Aberdeen,
Chemistry Department, Meston Walk, Aberdeen, AB24 3UE, Scotland}
\author{Claire H\'erold}%
\author{P. Lagrange}%
\affiliation{Institut Jean Lamour - UMR 7198 CNRS -
Nancy-Universit\'e - UPV-Metz -
D\'epartement Chimie et Physique des Solides et des Surfaces -
Facult\'e des Sciences et Techniques, B.P. 70239 - 54506
Vandoeuvre-l\`es-Nancy Cedex- France }

\author{Daniel Petitgrand}%
\affiliation{Laboratoire L\'eon Brillouin, CEA-CNRS, CE-Saclay, 91191
  Gif sur Yvette, France}

\author{Moritz Hoesch}%
\altaffiliation{present address: Diamond Light Source, Diamond House,
  Harwell Science and Innovation Campus, Didcot, Oxfordshire OX11 0DE, England}
\affiliation{European Synchrotron Radiation Facility, BP 220, F-38043
  Grenoble cedex, France}

\date{\today}

\begin{abstract}
We report measurements of phonon dispersion in CaC$_6$ using inelastic
X-ray and neutron scattering.
We find good overall agreement, particularly in the 50 meV energy
region, between experimental data and
first-principles density-functional-theory calculations.
However, on the longitudinal dispersion along the $(1 1 1)$ axis
of the rhombohedral representation, 
we find an unexpected anti-crossing with an
additional longitudinal mode, at about 11 meV.
At a comparable energy, we observe also unexpected intensity on the
in-plane direction. 
These results resolve the previous incorrect assignment of a
longitudinal phonon mode to a transverse mode in the same energy range. 
By calculating the electron susceptibility from first principles we show
that this longitudinal excitation is unlikely to be due to a plasmon and
consequently can probably be due to defects or vacancies present in the
sample.

\end{abstract}


\maketitle

\section{\label{intro}Introduction}

Intercalation of foreign atoms in graphite can stabilize superconductivity
by introducing metal atoms between the layers, which allow both tuning of
the inter-layer spacing and charging of the graphite host.  For a long
time it was believed that the maximum critical temperature obtainable
at ambient pressure in graphite intercalation compound (GIC)
\cite{enoki} was less than 2 K.
The discovery of high temperature superconductivity
in two intercalated compounds: YbC$_6$\cite{Weller2005} and
CaC$_6$\cite{Weller2005,emery} with unprecedented high transition
temperatures, 6.5 K and 11.5 K respectively, has raised renewed
interest about the role of phonons in GICs.

The role of the in-plane and out-of-plane phonon modes in CaC$_6$ has
been controversial.
By using the density functional theory, it has been shown that
superconductivity in CaC$_6$ is due to an electron-phonon mechanism
\cite{calandra:237002,Boeri06,Sanna07}.
The electron-phonon coupling is mainly associated to Carbon vibrations
perpendicular to the planes of graphite (C$_z$), and Calcium vibrations
parallel to the graphite planes (Ca$_{xy}$). However specific heat
measurements suggest \cite{Mazin06} that the contribution of C$_z$
vibrations is even larger than what was predicted by density functional
theory (DFT).
Consequently, it would be desirable to measure the phonon dispersion
of CaC$_6$ in order to see if there is actually a disagreement between
theory and experimental data.

In this work we measure phonons in CaC$_6$ along the $(1 1 1)$ axis
of the rhombohedral representation using inelastic
X-ray and neutron scattering (IXS and INS respectively). Furthermore,
in order to asses the precision of DFT simulations we compare the
calculated and measured in-plane averaged IXS structure factor.

On the longitudinal dispersion along the   $(1 1 1)$ axis
of the rhombohedral representation, we find an unexpected anti-crossing with an
unidentified additional mode, at about 11 meV.
At the same energy, we observe also unexpected intensity on the
in-plane direction. 

Previous measurements of phonon dispersion in CaC$_6$\cite{upton} assign
a longitudinal character to a mode in the same energy range, but the
mode is transverse. Here we show that and additional mode
appear in the same energy region, and interact with the longitudinal
acoustic mode.
This result could help to resolve this apparently incorrect mode
assignment. 

\begin{figure}
\includegraphics[scale=0.4,angle=0]{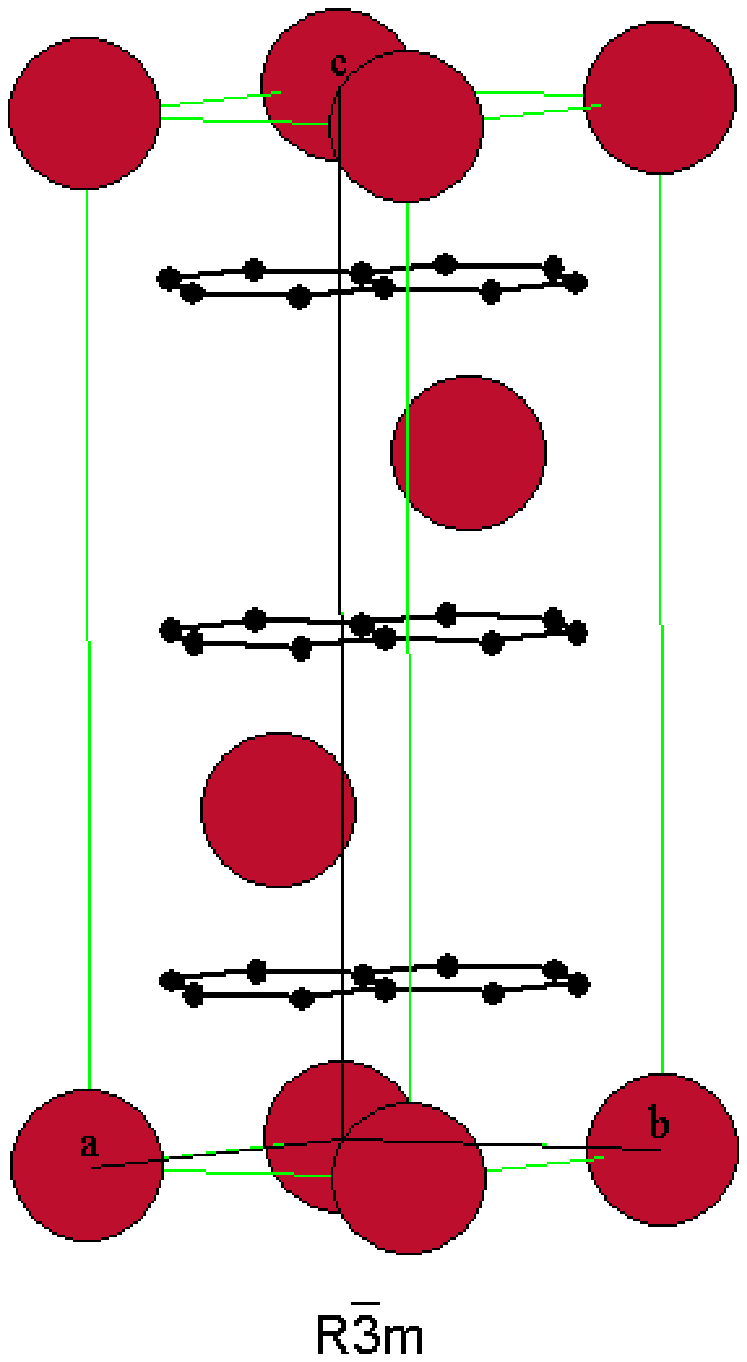}\hspace{48pt}\includegraphics[scale=0.4,angle=0]{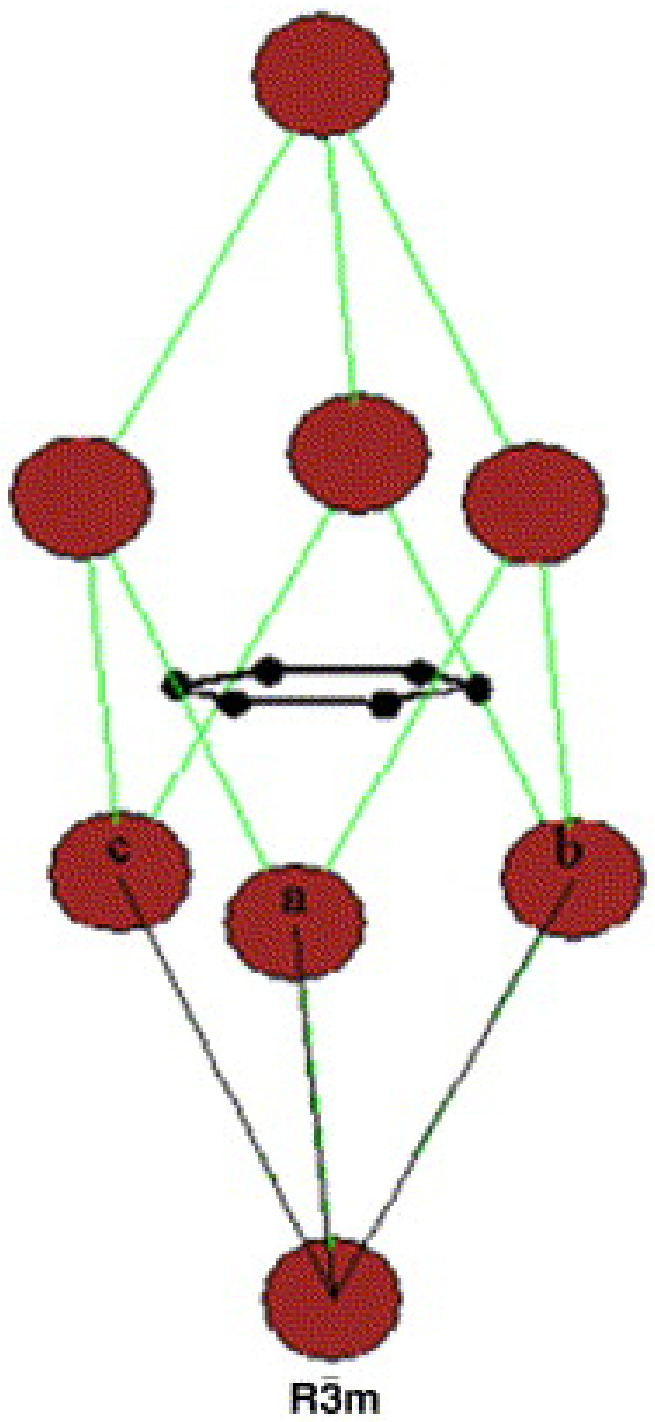}
\caption{\label{cryst} (Color online)
Right panel: rhombohedral crystal structure of CaC$_6$.
Large red circles correspond to calcium ions, small black ones to carbon atoms.
Left panel: The corresponding hexagonal unit cell.
The 3- rhombohedral c-axis is perpendicular to the honeycomb graphene planes.
}
\end{figure}

\section{\label{Method}Method}

IXS and INS measurements have been conducted on a polycrystalline sample of
highly oriented pyrolytic graphite intercalated with Ca,
prepared by liquid-solid synthesis method in lithium-based alloys
\cite{pruvost}.
The sample shows same superconducting properties \cite{sample-tc} as
previously reported on samples prepared using the same 
method \cite{emery}.   

Phonon measurements in graphite are a classical example of
comparison between IXS and INS, as shown at the very beginning
of the IXS technique by E. Burkel \cite{burkel}.
The IXS approach is particularly favorable for
high energy modes as it is the case in graphite and GICs where some
optical branches involving carbon ions are in the 170 meV energy range \cite{maultzsch}.

\begin{figure}
\includegraphics[scale=0.4,angle=0]{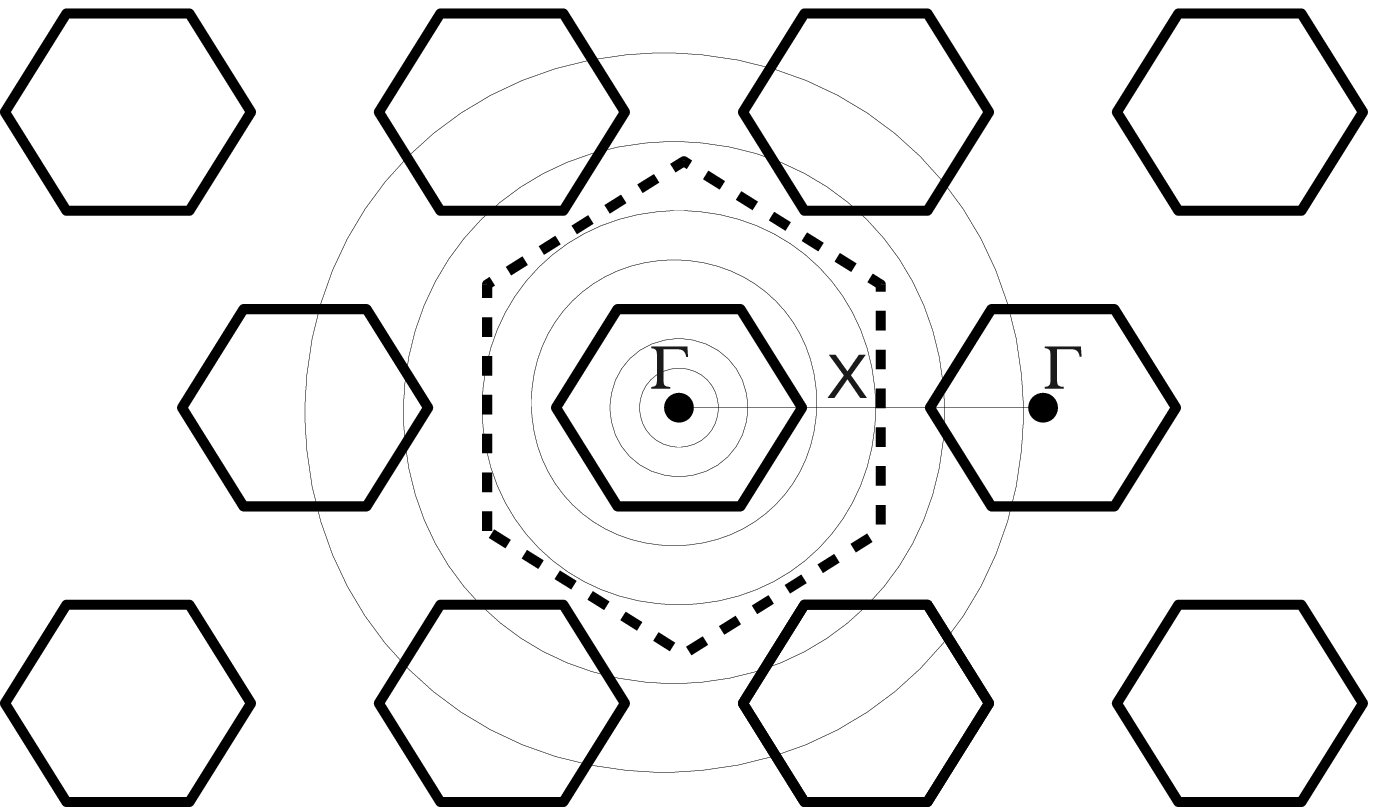}
\vspace{1pt}
\includegraphics[scale=0.4,angle=0]{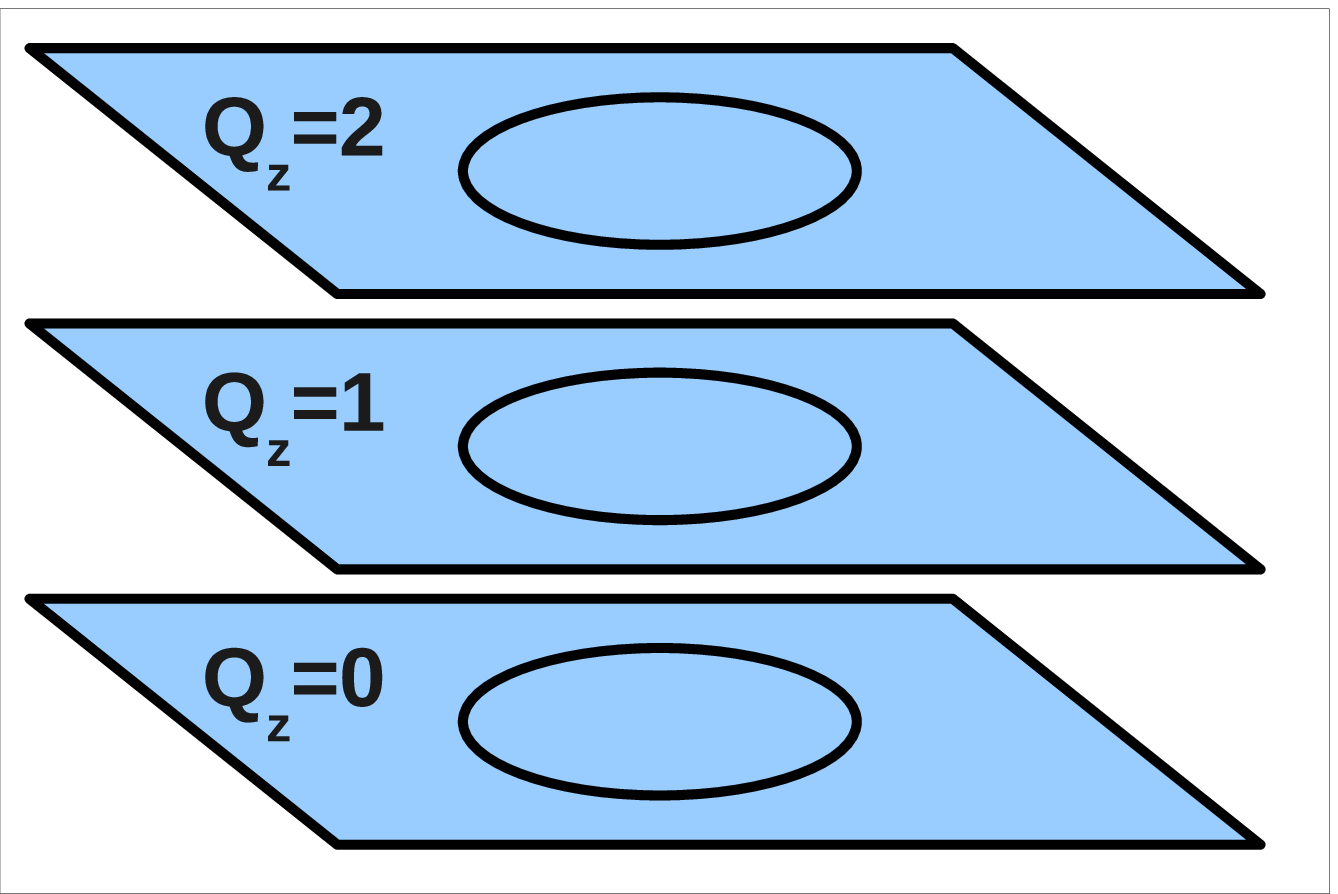}
\caption{\label{q-cut} (Color online) Top panel: section of the
  rhombohedral reciprocal space corresponding to the $a^*b^*$ plane of the
  graphite sub- lattice in CaC$_6$. Circle corresponds to fixed Q lines
  in the plane. Two $\Gamma$ points in different hexagons
  are shown. The line connecting the two $\Gamma$ points is parallel
  to $(1 -1 0)$. The large, dashed hexagon indicate the projection of the
  rhombohedral Brillouin Zone, with X the zone boundary along (1 -1 0).
  Bottom panel: the circles can be shifted in different
  Brillouin zones along the graphite c*, adding a component Q$_z
  \times$(1 1 1). The L point lies halfway between two planes, at
  Q$_z\times$(1 1 1)+(0.5 0.5 0.5).
}
\end{figure}

In this kind of sample the polycrystalline domains are actually
aligned along a crystal direction, corresponding to the 
the $(1 1 1)$ axis in the rhombohedral
structure of CaC$_6$ \cite{emery-stru} (see Fig.\ref{cryst}).
As a consequence, the phonon dispersion of the
longitudinal modes, with momentum parallel to the $(1 1 1)$
direction, can be measured.
The different Carbon layers in the sample are randomly rotated respect
to the common $(1 1 1)$ axis so that the IXS measurement for a given momentum
of modulus Q in the $a^*b^*$ plane will be an average over a circle of
radius Q, as shown in Fig. \ref{q-cut}, top panel.
The volume of the sample we measured was enough for a
measurement of the longitudinal phonon dispersion along (1 1 1) using
INS, but the signal from the in-plane
average was way too weak for neutron, while strong enough for
IXS. Therefore we decided to couple the
two probes in order to achieve an extensive insight on the phonon
structure of CaC$_6$.

The IXS experiment was carried out on the undulator ID28 beam-line at the
ESRF.
We have chosen to work with the Si (9 9 9) \cite{mono,mono2} reflection, with a
wave-length of 0.6968 \AA$^{-1}$ (17794 eV) and an energy resolution
$\Delta$E = 3.0 $\pm$ 0.2 meV \cite{ana}.
Additional spectra were collected
using the Si (11 11 11) reflection, with a wave-length of 0.5701
\AA$^{-1}$ (21747 eV) and an energy resolution $\Delta$E = 1.5 $\pm$
0.1 meV. The back-scattered beam is focused on the sample position
by a gold-coated toroidal mirror, which provides a focal spot of h
$\times$ v = 0.270 $\times$ 0.090 mm FWHM. Further details of the used
configuration are described in Ref. \onlinecite{mgb2}.
The sample measured in IXS is a platelet with thickness of $\sim$ 0.25 mm
along (1 1 1) direction and an area of 4.2 $\times$ 0.9 mm. 
The sample was hold in a Lindemann glass capillary, sealed in a glove
box.

\begin{figure}
\includegraphics[scale=0.6,angle=0]{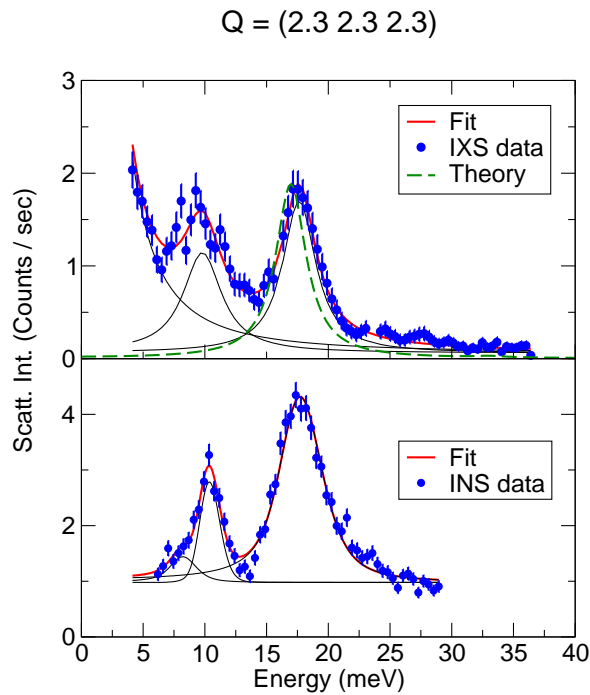}
\caption{\label{compXN} (Color online) Inelastic 
X-ray (top panel) and neutron (bottom panel) 
energy loss spectra at ${\bf Q} = {\bf G} +
  {\bf q}$ = 2$\times$ (1 , 1 , 1) + (0.3 , 0.3 , 0.3), corresponding
  to a longitudinal phonon polarization along 0.6 of the $\Gamma-L$ line.
  Data are fitted using a convolution of the instrumental function with
  an harmonic oscillator. In the top panel, theoretical calculations have
  been included (dashed lines).
}
\end{figure}

The inelastic neutron-scattering experiment was carried
out on the cold source 4F2 and thermal source 2T triple-axis
spectrometer at the Laboratoire L\'eon Brillouin in Saclay, France.
Harmonic contribution from the monochromator was reduced using a
standard graphite filter, so final neutron wave-number was typically of
1.975 \AA$^{-1}$ for 4F2, and of 2.662 \AA$^{-1}$ or 4.100 \AA$^{-1}$ for 2T
although some scan at 1.975 \AA$^{-1}$ was taken for
comparison sake or for averting artifacts from monochromator
harmonics.
Collimation were of 60'-open-sample-open-open,
and with a horizontally focusing monochromator and
a vertically focusing analyzer.
Several platelets, of about 6 mm width and for a total of 2 mm
thickness, were kept together in an aluminum foil. The sample was
sealed in a aluminum can, sealed in a glove box, using an Indium
gasket.

A comparison of the IXS and INS data is shown for a point along the
$\Gamma-L$ line in Fig. \ref{compXN}.

In parallel to IXS and INS data we carried out first-principles density
functional theory calculations in the linear response \cite{QE-2009, Baroni2001}.
We use the generalized gradient approximation \cite{PBE} and ultrasoft
pseudo-potentials \cite{Ultrasoft}. Technical details are the same
as in refs. \cite{calandra:237002,calandra:094507}.

\section{\label{ExpDis}Experiments and discussion}

\begin{figure}
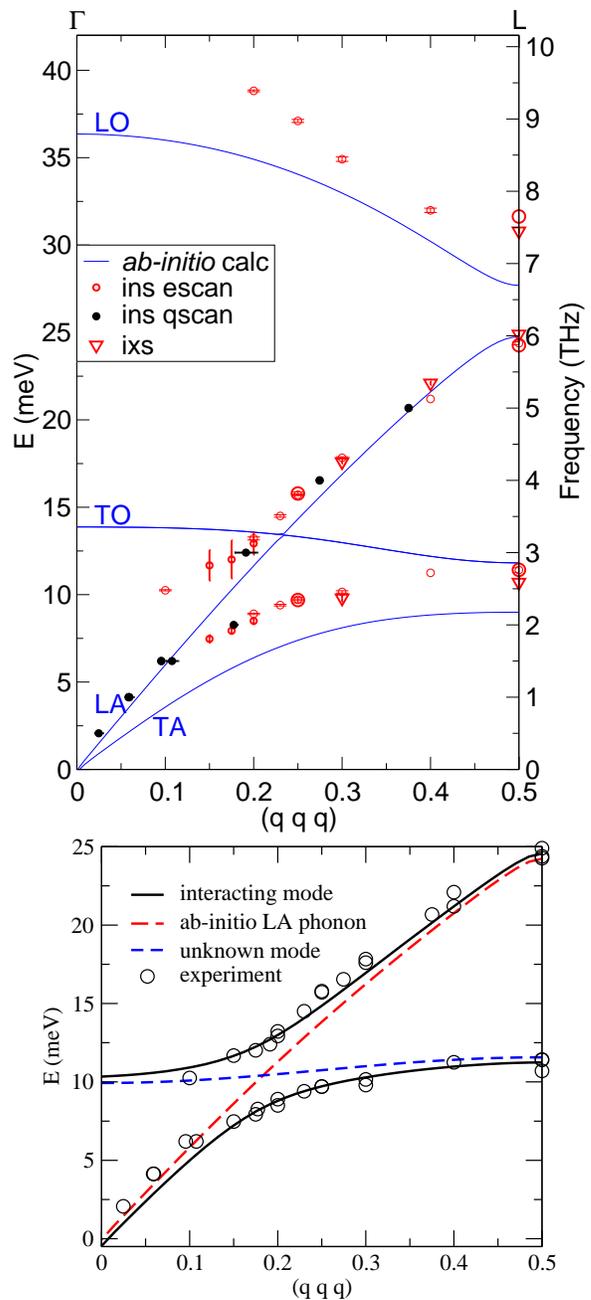

\includegraphics[scale=0.4,angle=0]{disp111-meV-pap.eps}
\includegraphics[scale=0.5,angle=0]{the_exp.eps}
\caption{\label{cdisp} (Color online)
Top panel: Phonon dispersion along the (1 1 1) direction. Hollow and
filled circles indicate, respectively, constant Q and constant energy
INS scan. Triangles represent IXS constant Q scan.
Lines represents ab-initio calculated
phonon dispersion. The scans are of type
$\mathrm{Q_z} \times$ (1 1 1) + (q q q) with
and $\mathrm{Q_z}$=1,2,3, in rhombohedral lattice.
Using standard notation, we label LA and TA the longitudinal and
transverse acoustic modes respectively. We label LO and TO
longitudinal and transverse optical modes, respectively.
Bottom panel: Zoom on the low energy part of the above dispersion data.
Circles indicate the experimental points (IXS and INS together).
Dashed lines show the dispersion of
the ab-initio longitudinal acoustic phonon and of the unknown mode.
Continuous lines show the dispersion of the interacting mode described
by Eq.~(\ref{eqmodel}), fitted to the experimental data.
}
\end{figure}

Figure \ref{cdisp}, top panel, shows the measured phonon dispersion along the
c-axis using IXS and INS, as compared with first principles calculations.
The low-energy phonons dispersion along (1 1 1) is generally in agreement
with theoretical calculations, with two notable exceptions.
First, the longitudinal optical mode is hardened in experiments
(the hardening is 7 meV at zone border).
Furthermore, although the sound velocity of the longitudinal
acoustic mode is in good agreement with DFT calculations,
above q = 0.1 and about 8 meV,
the longitudinal acoustic branch (LA) suddenly bends,
and its energy at the zone boundary is 11.4 $\pm$ 0.1 meV,
according to INS, which is very close to that
of the calculated first transverse optical mode at the zone boundary L
(labeled TO in Fig.\ref{cdisp}, top panel), at 11.8 meV.
The same behavior can be observed on the high energy section of the
LA mode.
Close to the zone boundary, from q = 0.5 to q = 0.3,
the dispersion, as measured with both INS and IXS, reproduces very well
the calculated one. Then it departs from the theoretical prediction
going towards the zone center, below q = 0.3 and 16 meV. Finally
towards zone center the LA mode flattens to a value of $\approx 10$
meV, slightly lower than what observed at L. 

Note, that the calculated energy for the TO mode at zone
center $\Gamma$ is substantially higher, namely $13.9$ meV.

\begin{figure}
\includegraphics[scale=0.6,angle=0]{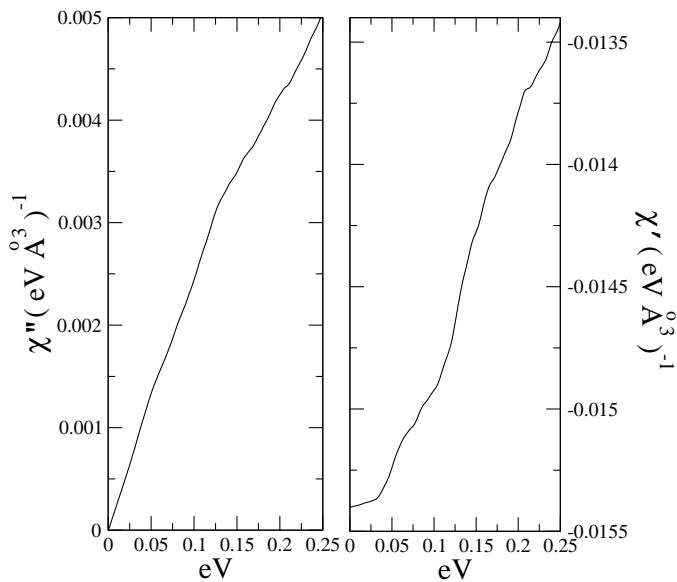}
\caption{\label{chi} Real and Imaginary part of the susceptibility
for CaC$_6$.}
\end{figure}

This behavior is reminiscent of an avoided crossing, or
anti-crossing, where the longitudinal acoustic mode interacts with an
unknown mode. To analyze the experimental data, we suppose that
the dispersion of the unknown mode is described by:

\begin{equation}
\epsilon(q)=a+b\cos(2\pi q)
\end{equation}
where $a$ and $b$ are fitting parameters and ($q$ $q$ $q$), is the momentum
in the rhombohedral coordinates.
We consider a $q$-independent coupling $\Delta$ with the longitudinal
acoustic phonon. Within this model, the excitation energies, observed
in the inelastic experiments, are the eigenvalues of a 2$\times$2 matrix:
\begin{equation}
\left( \begin{array}{cc}
\epsilon(q) & \Delta \\
\Delta & \hbar\omega_{\rm LA}(q) \end{array} \right),\label{eqmodel}
\end{equation}
where $\hbar\omega_{\rm LA}(q)$ is the longitudinal
acoustic phonon energy, as calculated using DFT.
By minimizing the mean square error between the computed and observed
energies, we obtain a good fit with $a=10.75$ meV, $b= -0.82$ meV and
$\Delta=2.06$ meV, see Fig. \ref{cdisp} (bottom panel).
As a consequence the detected anomaly can
be explained by the coupling of the longitudinal acoustic phonon mode
with a longitudinal excitation or defect mode of unknown origin.

\begin{figure*}
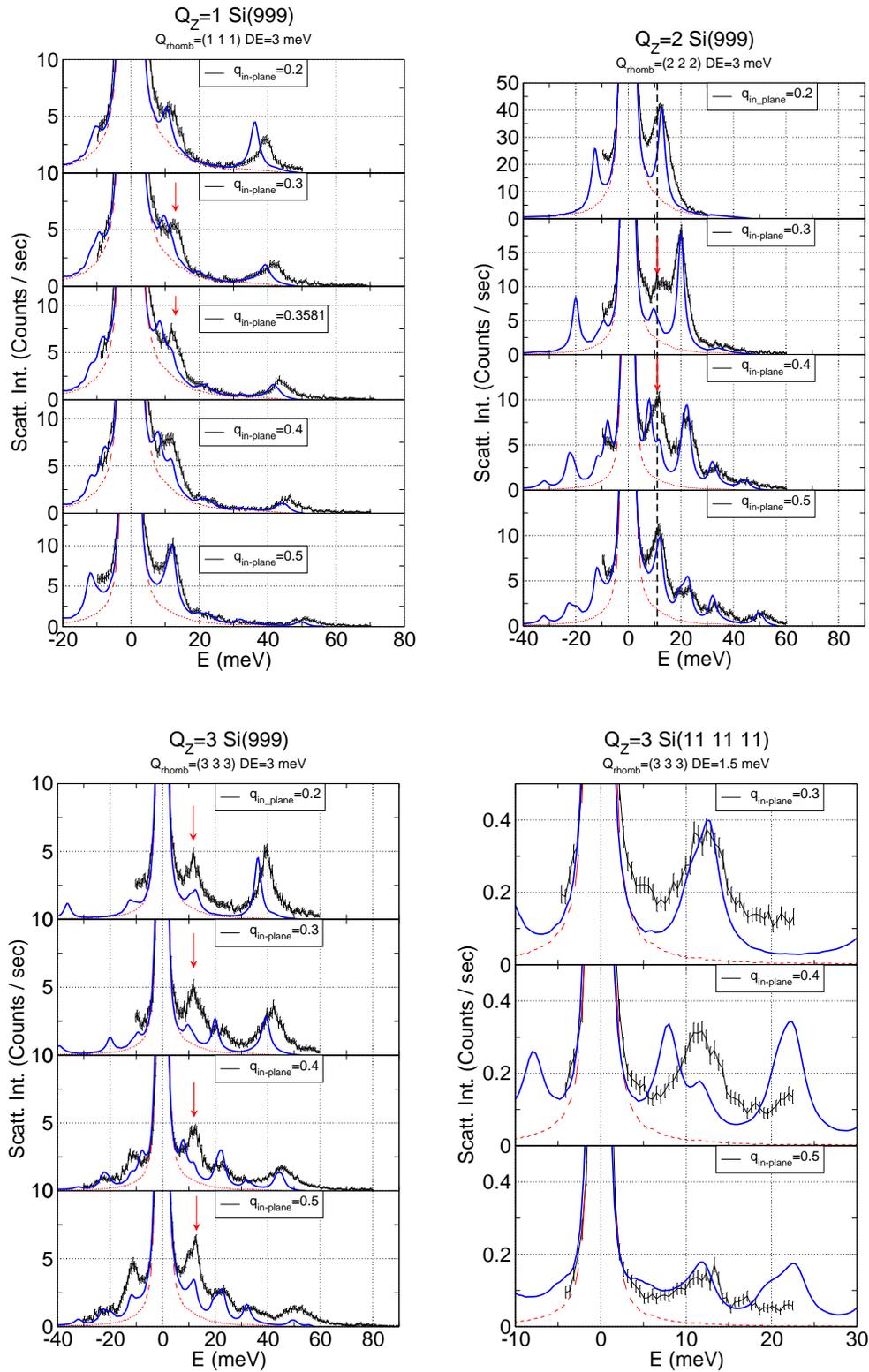

\includegraphics[scale=0.35,angle=0]{Q1Si9a2-fg.eps}\hspace{24pt}\includegraphics[scale=0.35,angle=0]{Q2Si9a2-f.eps}\vspace{24pt}
\includegraphics[scale=0.35,angle=0]{Q3Si9a2-f.eps}\hspace{24pt}\includegraphics[scale=0.35,angle=0]{Q3Si11a2-f.eps}
\caption{\label{abPDOS} (Color online)
Constant-Q in-plane phonon density of states
obtained using IXS (line with error bars) compared to
\textit{ab-initio} calculations (thick lines). The experimental
elastic contribution is also shown (thin lines).
The scans are obtained at ${\bf G}+ {\bf q}$ where
$\bf{G}=\mathrm{Q_z} \times (1 1 1)$ and $\mathrm{Q_z}$=1 (top, left),
z=2 (top, right), z=3 (bottom), while $\bf{q}$ lies in-plane and we
give its modulus along the (1 -1 0) direction in reduced length units
for each spectra.
For both top panels the energy resolution is 3 meV.
For the bottom panels the  energy resolution set-up is 3meV (left) and
1.5 meV (right).
Calculations are normalized to the mode at about 50 meV in the left panel,
while normalization is to the lowest energy mode in the right panel.}
\end{figure*}

The first optical mode is transverse
supposing R$\overline{3}$m symmetry, and therefore silent in this
configuration, as it can be seen in the simulated spectra in
Fig. \ref{compXN}. 
An interpretation of the anticrossing could be that the aforementioned
first optical mode would be activated by some symmetry breaking
scattering, as from impurities, which would mix different
polarizations.
However, the dispersion, flat within error bars, or
even with a slight positive slope at zone center, 
is not compatible with the negative slope of the
first optic.

A second possibility, could be the coupling to a 
longitudinal low energy plasmon.
Indeed in another layered material, MgB$_2$, an acoustic plasmon
mode was indeed detected at surprisingly low energy \cite{Silkin:054521}.
To address this issue we calculate the ${\bf G=0}$ and ${\bf G^{'}=0}$
bare susceptibility, namely
\begin{eqnarray}\label{eq:chi}
\chi_{\bf q}(\omega)&=&\chi_{{\bf q}}({\bf G=0},{\bf G^{'}=0},\omega)=\frac{1}{N_{\bf k} \Omega}\times \nonumber \\
& &\sum_{{\bf k},nm}
\frac{|\langle \psi_{{\bf k} n}|e^{-i{\bf q}\cdot{\bf r}}|\psi_{{\bf k+q} n'}\rangle|^{2}
(f_{{\bf k}n}-f_{{\bf k+q}m})}{\epsilon_{{\bf k}n}-\epsilon_{{\bf k+q}m}+\omega-i\eta}
\nonumber \\
\end{eqnarray}
where $\Omega$ =496.38 $ a_0^3$ is the unit cell volume with
$a_0$ = $\frac{\hbar ^2}{m e^2}$ = 0.53 \AA~ is the Bohr radius,
and $f_{{\bf k}n}$ is the Fermi function for a band energy $\epsilon_{{\bf k}n}$.
Since the calculation of Eq. \ref{eq:chi} requires a very accurate
sampling of the Brillouin
zone, we calculate the matrix-element using first-principles calculations on
a $6\times 6\times 6$ grid and then we interpolate it all over the
Brillouin zone using Wannier interpolation
\cite{Wannier90,Souza2002,MarzariV1997}.
In order to obtain a good description of the CaC$_6$
first principles bands, we use 7 Wannier functions.
Then we interpolate the matrix element and the band structure
over a $N_k=150\times 150\times 300$ k-points grid and calculate
$ \chi_{\bf q}(\omega)$ at $q=(0.25,0.25,0.25)$, close to where the
anti-crossing occurs.
The temperature in  the Fermi function is 
$T=300 K$ and the lorentzian smearing  $\eta=4 {\rm meV}$.\
The results for the real ($\chi^{'}$) and imaginary ($\chi^{''}$) part of the
susceptibility are shown in Fig. \ref{chi}.

A peak in the imaginary part of $\chi_{\bf q}(\omega)$ would signal
the occurrence of a plasmon excitation.
In our case the imaginary part of the susceptibility is featureless. 
Thus no plasmon excitations are present in CaC$_6$ calculated
susceptibility. 
To cross-check our result we also consider the case of MgB$_2$ where
a plasmon was found \cite{Silkin:054521}. In the MgB$_2$ case we
reproduce the occurrence  of an acoustic plasmon mode in agreement to
Ref. \onlinecite{Silkin:054521}.
As a consequence the 11 meV excitations is unlikely to be due to
a longitudinal plasmon, but most likely to impurities.

A possible origin of the impurity mode would be the presence of lithium in
the preparation of the samples by liquid-solid synthesis
method \cite{pruvost}.
This hypothesis is supported by a
strong XPS signal for the lithium ion at the K-edge (results not shown).
However, this measurements are not confirmed by nuclear microprobe
analysis.

A further possibility would be the presence of a P6$_3$/mmc phase
in the sample coexisting with the most stable R$\overline{3}$m one.
However the attempt to describe the c-axis phonon
dispersion in terms of a similar P6$_3$/mmc phase is also
unsatisfactory since the dispersion of the low energy modes is hardly
distinguishable from the dispersion of the $R\overline{3}m$  phase.
Still it is possible that the avoided crossing is generated by the
presence of vacancies or other similar defects.

An anomaly at a similar energy is also observed in another set of data
which consists in constant-Q 2D-phonon density-of-states, 
with a propagation vector ${\bf q}$ lying in the plane perpendicular to
the (1 1 1) direction and corresponding to the $a^*b^*$ plane.
For this IXS experiment, the choice of
${\bf G}$ = $\mathrm{Q_z} \times$ (1,1,1), with
$\mathrm{Q_z} \neq$ 0 resulted in a significant contribution from
transverse modes.
In this respect what is measured is a 2-dimensional
density of states obtained averaging over all the structure
factors having modulus of the IXS exchanged momentum
$| {\bf G}+ {\bf q} |$, where $| {\bf q} |$
is in the plane as reported in Fig.\ref{q-cut} and \ref{abPDOS}.
Even if the measurement is not equivalent
to a phonon dispersion calculation, as both longitudinal
and transverse modes are measured, the low energy
modes are fairly well reproduced, except for an extra intensity measured
at about 10 to 13 meV for all $\mathrm{Q_z}$, in particular for the
spectra with q from  0.3 to 0.4, (see arrows in Fig.\ref{abPDOS})
in agreement to what we found along the c-axis.

In a recent paper on a comparable sample of CaC$_6$ \cite{upton},
using IXS only, Upton and co-workers incorrectly assigned the flat
band at $\sim$11 meV to the first transverse optical mode.
This is due to the fact that they lack both a symmetry analysis of
their calculated modes as well as the resolution necessary to see the
details of the anti-crossing.
Note that the region close to zone center, below q = 0.15,
is difficult to measure using IXS due to the large elastic peak at
zero energy generated by disorder.
This is not the case in INS scattering data and only by using this
experimental technique the avoided crossing behavior between the LA
mode and an unknown mode can be revealed.

We note that for the first time, using the in-plane configuration with
IXS, we found good agreement between the data and the simulation in
the 50 meV energy region. This is particularly important as
vibrations propagating in-plane in this energy window are supposed to
be among the ones contributing to electron-phonon mechanism
\cite{calandra:237002, sugawara}.

\section{\label{Conclu}Conclusions}

In conclusion, we show that low energy phonon structure of the
superconducting graphite intercalated compound CaC$_6$, show clear
anomalies, in disagreement with DFT calculations assuming the perfect crystal
structure. In particular we observe an anti-crossing of the acoustic
longitudinal mode with an additional longitudinal flat mode.
Calculation of the electron susceptibility suggests that this mode
can not be attribute to a plasmon excitation.
Thus we infer that the unknown excitation is most
likely due to defects, impurities or vacancies.
On the other hand, we found good overall agreement between the data and the
simulations, particularly
in the 50 meV energy region where the C$_z$ vibrations
propagating in-plane in this energy window are supposed to provide
the largest contribution to electron-phonon mechanism.

\begin{acknowledgments}
M. D., M. C., G. L. and F. M. enjoyed fruitful discussion with A. C. Walters,
C. A. Howard, M. Ellerby, T.E. Weller, M.P.M Dean and S.S. Saxena.
We acknowledge M. Krisch for useful discussion and D. Gambetti for
technical help.
Calculations were performed at the IDRIS supercomputing center (project
081202).
S. Y. Z., J. G. were supported by the Director, Office of Science, Office
of Basic Energy Sciences, Materials Sciences and Engineering Division,
of the U.S. Department of Energy under Contract No. DE-AC02-05CH11231.
We acknowledge the support from University of California, Berkeley,
through France Berkeley Fund Grant, for reciprocal visit of the
Berkley and Paris team, to perform experiment, data analysis and
discussion.
This work was supported by ESRF through Experiment
No. HS-3189 and by LLB through Experiment No. 8170.
\end{acknowledgments}


\end{document}